\documentclass[prl,floatfix,twocolumn]{revtex4}
\usepackage{graphicx}
\usepackage{psfrag}

\def\eqn{earthquake}

\begin{document}

\title{Comment on ``Universal Distribution of Interearthquake
Times''}

\author{
J. {\AA}str\"om $^2$,
P.C.F.~Di Stefano$^{4
}$,
F.~Pr\"obst$^1$,
L.~Stodolsky$^{1 *}$,
J.~Timonen$^3$,
}

\affiliation{
 $^1$ Max-Planck-Institut f\"ur Physik, F\"ohringer Ring
6, D-80805 Munich, Germany;
$^2$  CSC - IT Center for Science, P.O.Box 405, FIN-02101 Esbo, 
Finland;
$^3$ Department of Physics, P.O. Box 35 (YFL), FIN-40014 University
of Jyv\"askyl\"a, Finland;
$^4$  Institut de Physique Nucl\'eaire de Lyon, Universit\'e Claude
Bernard Lyon
I, 4 rue Enrico Fermi, 69622 Villeurbanne Cedex, France;
$^*$ Corresponding author, {\it email address:}  les@mppmu.mpg.de.}

\begin{abstract}
\end{abstract}

\maketitle

 In a  {\it Letter} earlier this year ~\cite{sorn} and in a
number of  preceeding publications~\cite{cor}\cite{sca}\cite{bak}, 
 the  probability
distributions for the ``waiting time'' between earthquake events 
have been discussed. In particular it appears that 
the probability distribution for the number of events with waiting
time $w$, when expressed in terms of a suitably scaled
variable  $(w/w_0)$ with $w_0$  some characteristic time constant, 
follows a universal
function~\cite{bak}. In this {\it Comment} we would like to draw
attention to the
fact that recently published data \cite{crst} of the CRESST
collaboration on microfractures  in sapphire show the same
features.  Indeed there is  a great similarity, if not a
remarkable complete identity, of the probability distributions
expressed in this manner between the \eqn s and the microfractures.

\begin{figure}[h]\label{ww}
{{\includegraphics[width=\hsize]
{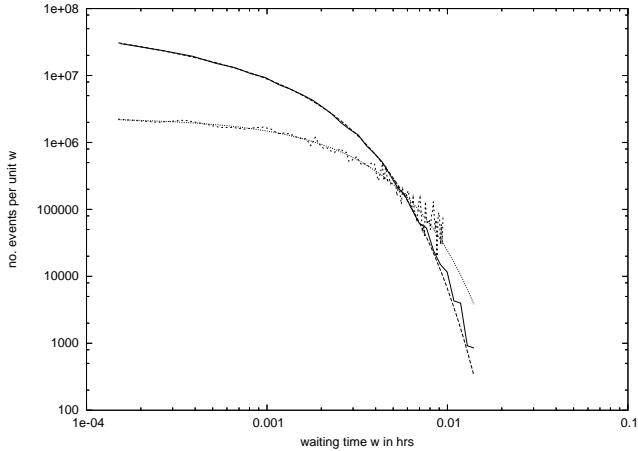}}}
\caption{ CRESST waiting time distributions. Upper curve:
microfractures,
  fit to $\propto w^{-\alpha}e^{-w/w_0}$. Lower curve:
 photon-induced events from  a calibration run,  fit to
$\propto e^{-w/w_0}$.}
\end{figure}

In Fig 1 we reproduce Fig 2 of ref~\cite{crst}. The upper curve is
the data on microfractures, fit to  
\begin{equation}\label{one}
 dN/dw\propto w^{-\alpha}e^{-w/w_0}
\end{equation}
 with $\alpha=0.33$ and $w_0=0.0014$ hrs. It will
be seen there
is an excellent fit.  The lower curve
represents a test of the
apparatus and  analysis, using photon-induced
events from an  external radioactive source. These should
follow the Poissonian $e^{-w/w_0}$ and there is also a good
fit.

According to Corral   ( Physica A)
the  form Eq~\ref{one} describes the waiting times for \eqn s, and
with  the same power, $\alpha=0.33$.   
 Concerning the time
scale parameter $w_0$, it is essentially the inverse of the
observational or experimental event rate R since from  Eq~\ref{one}

\begin{equation}\label{two}
1/R={\bar w}=(1-\alpha)w_0\;.
\end{equation}
 
The data used in the Figure satisfy this relation to within a few
percent, as would be expected from the good fit. We find that
raising the energy threshold in a data set, and so  reducing R,
leads to a linear relation between the fit $w_0$ and R, as would be
expected from  Eq~\ref{two} with a constant $\alpha$. 

 Alternatively one
could renounce fitting $w_0$, and simply  substitute
$w_0^{-1}=(1-\alpha)R$ into  Eq~\ref{one}, use the
experimental R (=28 000 events/28.5 hrs),   and fit for $\alpha$.
This essentially one parameter fit is satisfactory and yields
$\alpha=0.26$.

Although the CRESST values for $\alpha$ thus vary somewhat
according to the analysis and from run to run, the  parallelism
between the two kinds of phenomena  is striking. 
These considerations, involving such widely disparate time scales,
energies, and material properties, raise the question as to whether
Eqs \ref{one} and \ref{two} do not represent a general law,
applicable to many kinds of fracture processes.


\begin{thebibliography}{00}

\bibitem{sorn} A. Saichev and D. Sornette, Phys. Rev. Lett. {\bf
97}, 078501, (2006).

\bibitem{cor}   A. Corral, Phys Rev. {\bf E 68}, 035102
(2003); A. Corral, Physica  (Amsterdam) {\bf A 340}, 590 (2004);

\bibitem{sca} N. Scafetta and B. J. West, Phys. Rev. Lett. {\bf
92}, 138501, (2004).

\bibitem{bak}  P. Bak, K. Christensen, L. Danon, and T. Scanlon,
 Phys. Rev. Lett. {\bf 88}, 178501, (2002);

\bibitem{crst} J. {\AA}str\"om et al., Phys. Lett. {\bf A356} 262
(2006), 
(arXiv.org: physics/0504151); Nucl. Inst. Methods {\bf A559}, 754
(2006).


\end{thebibliography}
\end{document}